# Non-destructive splitter of twisted light

Yan Li,[1,2] Zhi-Yuan Zhou,[1,2,#] Dong-Sheng Ding,[1,2] Wei Zhang,[1,2] Shuai Shi,[1,2]

Bao-Sen Shi,[1,2*] and Guang-Can Guo[1,2]

[1]*Key Laboratory of Quantum Information, University of Science and Technology of China, Hefei,*

*Anhui 230026, China*

[2]*Synergetic Innovation Center of Quantum Information & Quantum Physics, University of Science*

*and Technology of China, Hefei, Anhui 230026, China*

[*] *drshi@ustc.edu.cn*
[#] *zyzhouphy@mail.ustc.edu.cn*

Efficiently discriminating beams carrying different orbital angular momentum (OAM) is of fundamental importance for various applications including high capacity optical communication and quantum information processing. We design and experimentally verify a distinguished method for effectively splitting different OAM-carried beams by introducing Dove prisms in a ring cavity. Because of rotational symmetry broken of two OAM-carried beams with opposite topological charges, their transmission spectra will split. When mode and impedance matches between the cavity and one OAM-carried beam are achieved, this beam will transmit through the cavity, and other beam will be reflected without being destroyed their spatial shapes. In this case, the cavity acts like a polarized beam splitter. The transmitting beam can be selected at your will. The splitting efficiency can reach unity if the cavity is lossless and it completely matches with the beam. Beams carry multi-OAMs can also be effectively split by cascading ring cavities.

The full degrees of freedoms of light beam include frequency, intensity, polarization and orbital angular momentum (OAM). It has been shown that a beam with azimuthal phase of $e^{il\varphi}$ carries $l\hbar$ OAM [1]. Singularities in intensity and phase distributions of OAM-carried beams have stimulated many exciting applications such as optical manipulation and trapping [2, 3], high precision optical metrology [4-6] and quantum information processing [7-14]. Mutual-orthogonal OAM modes offers the possibility of spatial mode multiplexing for high capacity optical communications [15]. The unlimited dimensions of OAM modes also hold promising for dense coding [16, 17] and coordinate independent quantum key distribution [18].

For various applications based on multi-OAM modes, efficiently discriminating and separating different OAM modes are of fundamental importance. In history, there are many methods for this target. For example, a hologram grating and a single mode fiber can be used as a mode detector for a specific OAM mode [15, 19]. In this case, the mode detector is a projector, we need *N* projection measurements to measure *N* OAM modes and the original state is destroyed completely after measurement. The projection measurements is not efficient, a more efficient method is to use wave front transformation from Cartesian to log-polar coordinate [20]. After transformation, the azimuthal phase profile of an OAM mode is mapped to a tilted planner wavefront. As a result, different OAM modes are mapped to different transverse tilts being proportional to their topological charges in the image plane, the separation efficiency is limited to theoretical limit of 77%. Then an improved scheme is demonstrated by using multi-copy of the unitary transformed refractive beams to enhance the separation efficiency to be 97% [21]. In addition to these methods, another method for separating OAM modes through spin-orbital coupling effect is proposed, which is based on Mach-Zehdner interferometer with Dove prisms in each arm [22-24]. This method can separate different OAM modes according to their parity, the separation efficiency can reach 100% in principle. To separate N OAM modes one needs N-1 cascaded interferometers. The difficulty in designing and aligning such a system limits its applications.

In this article, we propose and experimentally demonstrate a different OAM modes splitting scheme by introducing Dove prisms inside a ring cavity. The rotational symmetry for two OAM modes with opposite topological charges is broken by relatively rotating the prisms' axes, leading to non-degenerate of the cavity's transmission spectrum. One of the modes transmits when it is on resonance in cavity, the other mode is then reflected. This scheme has several advantages: 1. the two modes are split non-destructively after separation, therefore the splitter acts like a polarized beam splitter. 2. Our method is capable of splitting both symmetry and asymmetry superposition modes. 3. The transmitting mode can be selected at your will. 4. Beam carries multi-OAMs can also be effectively separated by cascading cavities. In the following, we first give numerical simulation of the performance of this device with cavity loss and impedance mismatch, then we show proof of principle experiments to give a glance of the performance of our device.

For a typical ring cavity, OAM modes are also eigenmodes of the cavity, the frequency of a certain OAM mode being resonant with the vacuum cavity can be expressed as

$$v_{plq} = \frac{c}{L}[q + \frac{1}{2}(l + 2p + 1)] \quad (1)$$

Where $c$ is the speed of light in vacuum, $L$ is the round trip optical length of the cavity, $q$ is a integer, $l$ is the toplogical charge, and $p$ is the radical index of OAM beam. The minimum frequency difference of a specific mode is $\Delta v_{plq} = c/L$. Two OAM modes with opposite topological charges are degenerate in the vacuum cavity. If we put two Dove prisms inside the cavity, there will be a relative phase $\Delta\Phi = 2l\alpha$ between the two OAM modes, where $\alpha$ is the relative rotation angle of two Dove prisms' axes. Then the two OAM modes will experience different phase change in one round trip, inducing their spectral splitting.

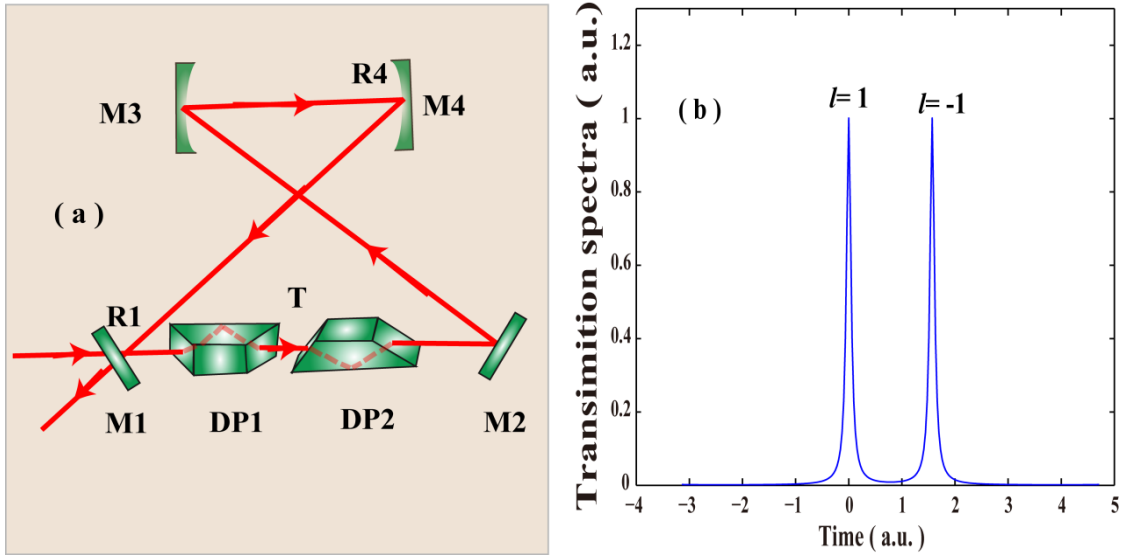

Figure 1. Non-degenerate cavity configuration and transmission spectra. (a) Ring cavity with two Dove prisms; (b) transmission spectra for two OAM modes with opposite topological charges, here we assume perfect mode and impedance matching of the cavity. M1-M4: mirrors; DP1-DP2: Dove prisms; R1,R4: reflectance of mirrors M1 and M4; T: internal cavity transmittance excludes losses of mirrors M1 and M4.

For the transmittance and reflectance parameters shown in figure 1(a), the transmission and reflection coefficient of the ring cavity can be expressed as following:

$$C_R = \frac{(\sqrt{R_1} - \sqrt{R_4 T})^2 + 4\sqrt{R_1 R_4 T}\sin^2(\delta/2)}{(1 - \sqrt{R_1 R_4 T})^2 + 4\sqrt{R_1 R_4 T}\sin^2(\delta/2)} \quad (2)$$

$$C_T = \frac{T_1 T_4 T}{(1 - \sqrt{R_1 R_4 T})^2 + 4\sqrt{R_1 R_4 T}\sin^2(\delta/2)} \quad (3)$$

Where $R_1$, $R_4$ and $T_1$, $T_4$ are the transmittance and reflectance of the cavity mirrors M1 and M4 respectively; $T$ represents the internal cavity losses including the loss of the Dove prisms and loss from other cavity mirrors; $\delta$ is the phase shift in one round trip.

We first numerically simulate the transmission and reflection coefficient (CT, CR) of the cavity for various cavity loss and impedance mismatch. The simulation results are shown in figure 2(a)-2(c). In figure 2(a), the internal cavity loss and the transmittance of the output mirror are fixed and the input mirror transmittance is an adjustable parameter. While in figure 2(b), the internal loss is an adjustable parameter, the transmittances of the input and output mirrors are fixed. We can see that CR reaches minimum but CT reaches the maximum for impedance matching case in figure 2(a) and 2(b). CT decreases while CR increases with the increase of impedance mismatching. In figure 2(c), we show the transmission coefficient for various cavity losses at the condition of impedance matching $T_1 = T_4 + T$. In this situation, CR keeps zero and CT decreases with the increase of internal loss. In figure 2(d), we simulate the transmission and reflection mode shapes for various phase shifts ΔΦ for input mode

$1/\sqrt{2}(|1\rangle+|-1\rangle)$ when the cavity is on resonance with mode $|1\rangle$. When $\Delta\Phi$ increases, the transmission spectra of modes $|1\rangle$ and $|-1\rangle$ become splitting from the degenerate states to non-degenerate states. When the transmission spectra of two modes are totally separated, the transmitting mode will only contain the mode which is resonant with cavity, the other mode is totally reflected. The reflection coefficient of mode $|-1\rangle$ will increase from 0 to 1. In figure 2(e), we simulate the reflection and transmission spatial shapes for various matching coefficient M ranging from 0 to 0.995 for input mode $1/\sqrt{2}(|1\rangle+|-1\rangle)$. For M=0, both modes are totally reflected, there will be no transmission of modes from the cavity; while for nearly perfect mode matching M=0.995, one of the mode is totally reflected and the other mode transmits completely.

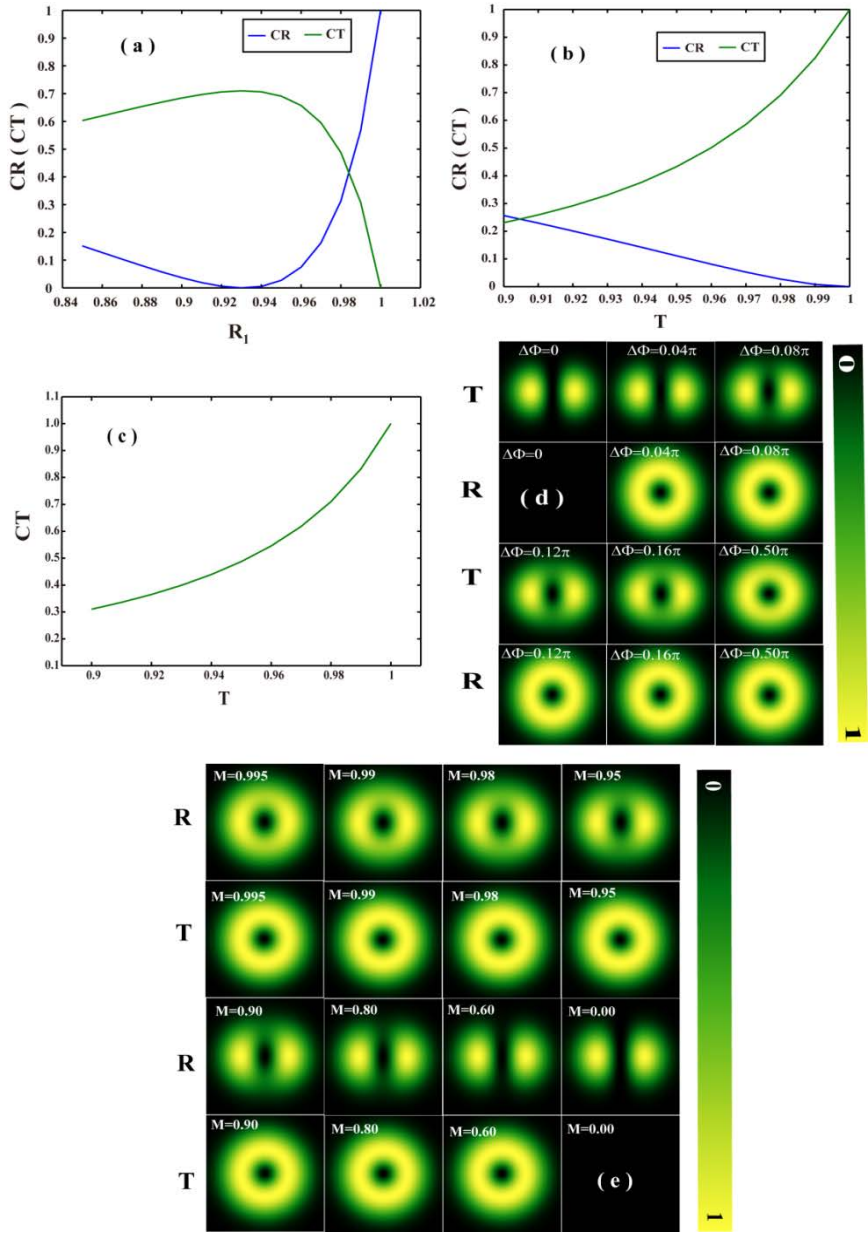

Figure 2. Simulation results for various cavity parameters. (a) CR and CT of the cavity as function of reflectance $R_1$, $R_4=0.95$, $T=0.98$; (b) CR and CT of the cavity as function of internal transmittance $T$, $R_1=R_4=0.95$; (c) CT as function of transmittance $T$ at the condition of impedance matching, $R_4=0.95$ in the calculation; (d) transmission and reflection spatial shapes of the cavity for various phase shift between modes $|1\rangle$ and $|-1\rangle$ for input mode $1/\sqrt{2}(|1\rangle+|-1\rangle)$, $R_1=R_4=0.90$, $T=1$ in the calculation; (e) transmission and reflection spatial shapes of the cavity for different mode matching efficiency $M$ ranges from 0.995 to 0, $R_1=R_4=0.90$, $T=1$ in the simulations.

Next, we perform proof of principle experiments for a few superposition modes. The experimental setup is depicted in figure 3. The 795 nm laser is from a continuous wave Ti: Sapphire laser (Coherent, MBR110), the polarization of the beam is controlled with quarter and half wave plates (QWP, HWP) before it is transformed to OAM-carried beam by a spatial light modulator (SLM); the OAM-carried beam is mode-matched to the ring cavity using lenses L1 and L2. The ring cavity consists of mirrors M1-M4. Two Dove prisms (DP1, DP2) are placed between planar mirrors M1 and M2, a piezo-transducer (PZT) is attached to mirror M2 for scanning and locking the cavity. Two concave mirrors M3 and M4 have radius of 80 mm. The input coupling mirror M1 has 85% reflectance at 795 nm, the output coupler mirror M4 has 95% reflectance at 795 nm, mirrors M2 and M3 are high reflection (R>99.9%) coated at 795 nm. The total length of the cavity is 600 mm. The transmitting beam from the cavity is imaged using CCD camera.

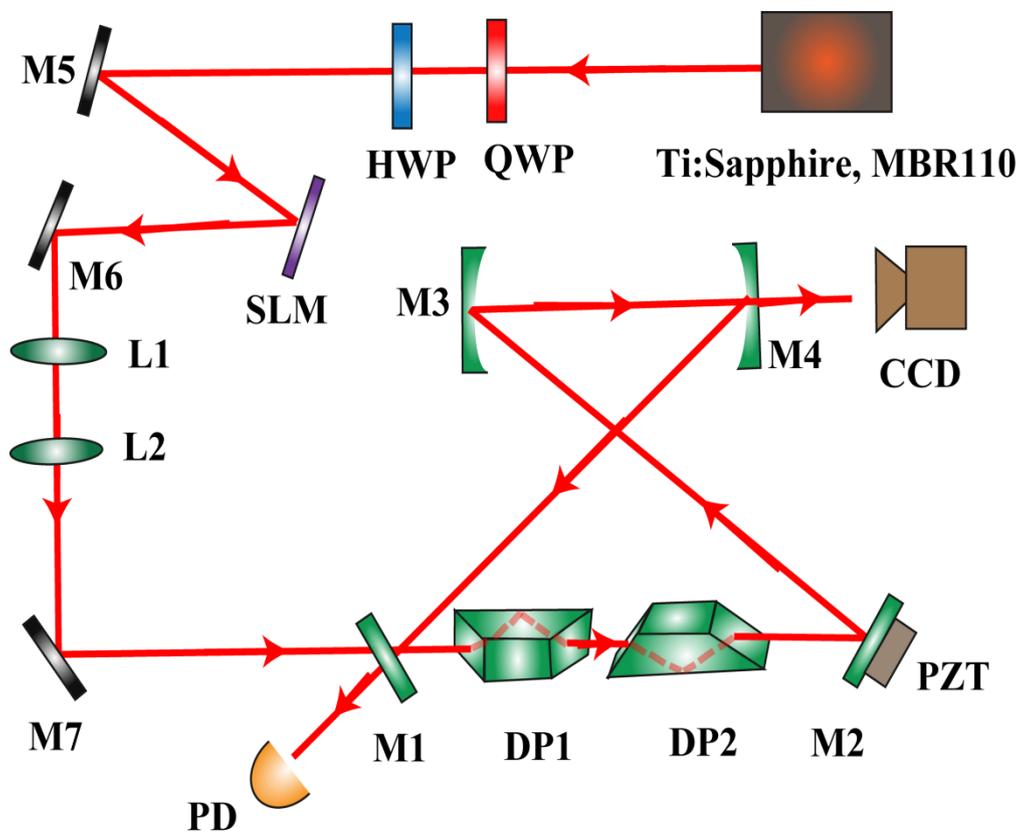

Figure 3. Experimental setup of our experiments. QWP(HWP): quarter (half) wave plate; M1-M7: mirrors; L1-L2: lenses; SLM: spatial light modulator; DP1, DP2: Dove prisms; PZT:

piezo-transducer; CCD; PD: photodiode; charge coupler device camera.

The experimental and simulation results for different input modes are shown in Figure 4. Figure 4(a)-4(d) are the transmission spectra for input modes of $|0\rangle+|1\rangle$, $|1\rangle+|2\rangle$, $|-1\rangle+|1\rangle$ and $|-2\rangle+|2\rangle$, respectively. The transmission spectra are detected using fast photodiode (PD) and acquired by an oscilloscope. For higher topological charge OAM modes, there will be more irrelevant higher cavity modes in the transmission spectra. The internal cavity loss for higher OAM mode is also higher, which leads to lower transmission efficiency. In our experiments, the internal cavity transmittance T=0.90, the total mode and impedance matching coefficient is about 0.80 to 0.90, which means there are about 0.10 to 0.20 ratio of the resonance mode is reflected. When the cavity is locked to one of input mode using Hasch-Coillaud (HC) method [25], the reflected and transmitted spatial shapes acquired using CCD camera are shown in left group of images in figure 4(e). We find that the transmitted beam has a well distinguished donut shape, while the reflected beam is distorted because of partially reflected resonant modes arising from mode mismatching. Another reason for poor reflected beam shape is the impurity of the input mode, as our SLM is act as a reflecting mirror, there should be other un-modulated part in the input beam, which will be reflected from the cavity and contaminated the reflected modes. Right group of images are the corresponding theoretical simulation results based experimental parameters.

In the present experiments, the internal cavity loss is about 10%, which is mainly from surface reflections of two Dove prisms. The transmission efficiency of the resonance mode is not very high. We are also prevented from testing higher topological charges because of higher modes losses and poor surface quality of the cavity mirrors, but these problems can be solved with the state-of –art mirror fabrication and coating technology. Another optimizing method is to minimize the size of the device and encapsulate it in a confined environment, then the cavity does not need active locking. If 95% transmission efficiency of the resonance mode is reached in the future, such device will act like a polarized beam splitter, which will be of great importance in many OAM-carried light based applications especially in optical communications [15-18], quantum simulations[26-28]. By cascading the basic device, one can achieve an OAM modes sorter for multi-OAM carrying beams. Our device has the evident advantage that OAM modes do not been destroyed after they are separated, another distinguished property of our device is that the transmitted mode can be selected at your will by locking the cavity to the mode on your demand.

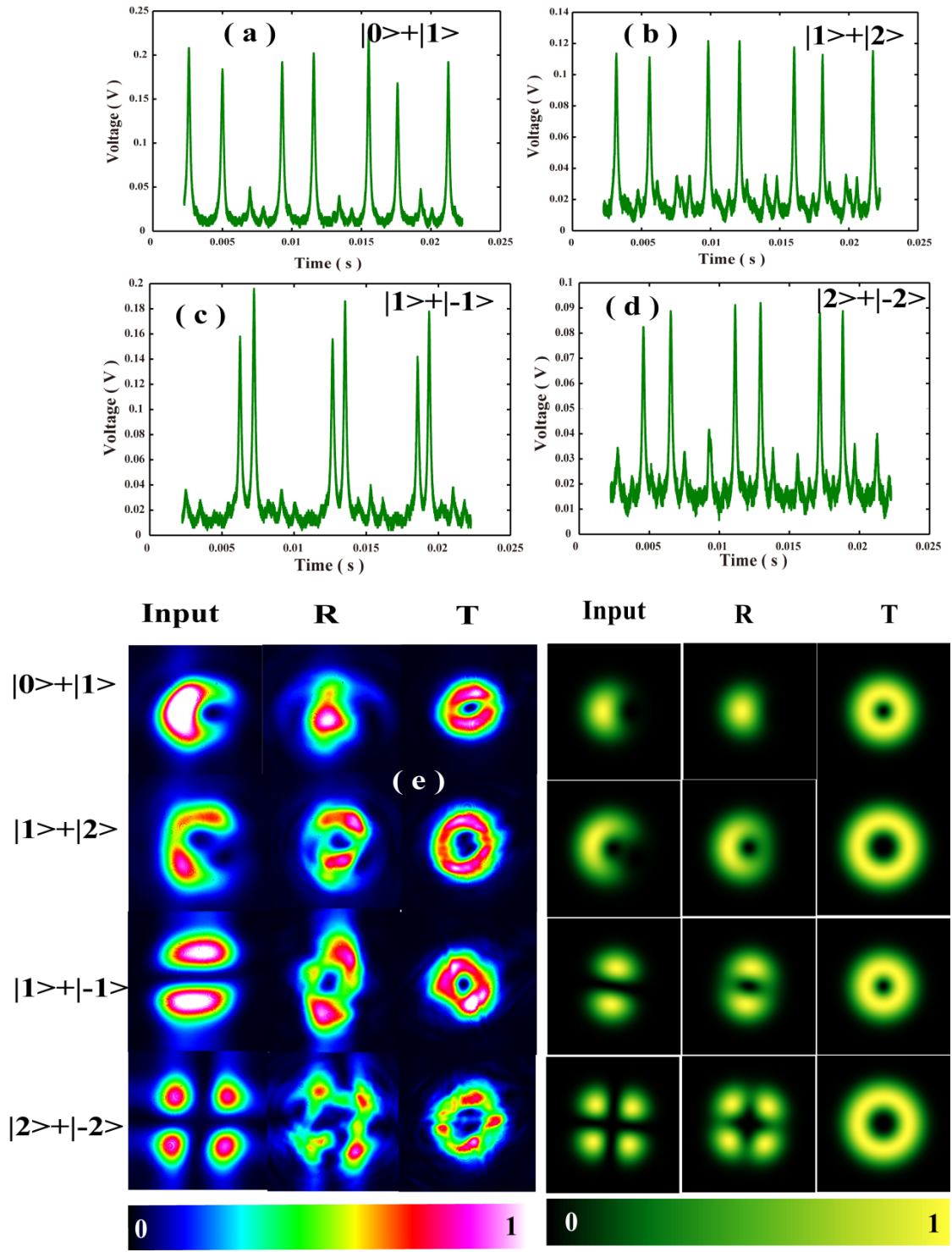

Figure 4. Experimental results for different modes input. (a)-(d) are transmission spectral for different input modes; (e) left group of images shows the intensity distribution for input mode, reflected mode and transmitted mode, right group of images are the corresponding numerical simulation images, the cavity is resonance for modes $|1\rangle, |2\rangle, |1\rangle$ and $|2\rangle$ for the four input superposition mode respectively.

In conclusion, we propose and experimentally verify a promising OAM-carried light beam splitter based on spatial symmetry broken in a ring cavity. The symmetry broken is introduced by two Dove prisms, which leads to non-degenerate cavity modes for two OAM-carried beams with opposite topological charges. Both theoretical simulations and experimental verification are performed to demonstrate its performances with different parameters. We believe that our device will be feasible to be broadly used in both classical and quantum regime for OAM-carried beam based optical communications and quantum information processing.

**Acknowledgements**
This work was supported by the National Fundamental Research Program of China (2011CBA00200), the National Natural Science Foundation of China (11174271, 61275115, and 61435011).